\documentclass[3p]{elsarticle}

\usepackage{hyperref}
\usepackage{amssymb}
\usepackage{bm}
\usepackage{dcolumn}
\usepackage{amssymb}
\usepackage{amsmath}
\usepackage{graphicx}
\usepackage{booktabs}
\usepackage{multirow}
\usepackage{setspace}
\usepackage{array}
\usepackage{threeparttable}
\usepackage{txfonts}
\journal{APP, published in Astropart. Phys. 82 (2016) 72 -- 76}

\begin{document}

\begin{frontmatter}



\newcommand*{\PKU}{School of Physics and State Key Laboratory of Nuclear Physics and
Technology, Peking University, Beijing 100871,
China}
\newcommand*{\CIC}{Collaborative Innovation Center of Quantum Matter, Beijing, China}
\newcommand*{\CHEP}{Center for High Energy Physics, Peking University, Beijing 100871, China}
\newcommand*{\CHPS}{Center for History and Philosophy of Science, Peking University, Beijing 100871,
China}

\title{Light speed variation from gamma-ray bursts}

\author[a]{Haowei Xu}
\author[a,b,c,d]{Bo-Qiang Ma\corref{cor1}}

\address[a]{\PKU}
\address[b]{\CIC}
\address[c]{\CHEP}
\address[d]{\CHPS}
\cortext[cor1]{Corresponding author \ead{mabq@pku.edu.cn}}

\begin{abstract}
The effect of quantum gravity can bring a tiny light speed variation which is detectable through energetic photons propagating from gamma ray bursts (GRBs) to an observer such as the space observatory. Through an analysis of the energetic photon data of the GRBs observed by the Fermi Gamma-ray Space Telescope (FGST), we reveal a surprising regularity
of the observed time lags between photons of different energies with respect to the Lorentz violation factor due to the light speed energy dependence. Such regularity suggests a linear form correction of the light speed $v(E)=c(1-E/E_{\rm LV})$, where $E$ is the photon energy and $E_{\rm LV}=(3.60 \pm 0.26) \times 10^{17}~ \rm GeV$ is the Lorentz violation scale measured by the energetic photon data of GRBs. The results support an energy dependence of the light speed in cosmological space.
\end{abstract}


\end{frontmatter}

It is a basic assumption in Einstein's relativity that the speed of light is a constant $c$ in free space. However, it is speculated from quantum gravity that the Lorentz invariance will be broken at the Planck scale ($E\sim E_{\rm P1} = \sqrt{\hbar c^5 / G}\approx 1.22 \times 10^{19} ~ \rm GeV$), thus the light speed may receive a correction of the order $E_{\rm photon}/E_{\rm Pl}$ due to the Lorentz invariance Violation (LV). As the photon energy $E_{\rm photon}$ is very small in comparison with the Planck energy $E_{\rm Pl}$, the correction to the light speed is too tiny to be detectable in most circumstances. Amelino-Camelia {\it et al.}~\cite{method1, method2} first suggested to use distant astrophysical sources of energetic photons to test the light speed energy dependence. For energetic photons %with very high energy ($>10~\mathrm{GeV}$)
from a gamma ray burst (GRB), the large cosmological distance from the source to an observer can amplify the tiny photon speed variation into observable quantities such as the arrival time lags between photons with different energies.
The Fermi Gamma-ray Space Telescope (FGST)~\cite{LAT, GBM} is a space observatory launched in 2008 to perform gamma-ray astronomy observations.
The Fermi Large Area Telescope (LAT) on board the FGST has detected a number of GRBs with over 10~GeV photons~\cite{F-LAT_above_10, data}. Provided that the redshifts of some GRBs have been measured by optical and X-ray telescopes, these data would offer us an opportunity to find evidence for or against the light speed variation in cosmological space.

For energy $E\ll E_{\rm Pl}$, the modified dispersion relation of the photon can be expressed in a general form as the leading term of Taylor series
\begin{equation}\label{eq:1}
  E^2=p^2 c^2 \left[1-s_n\left(\frac{pc}{E_{\mathrm{LV,} n}}\right)^n\right],
\end{equation}
from which we can derive the modified light speed, using the relation $v=\partial E/\partial p$,
\begin{equation}\label{eq:2}
  v(E)=c\left[1-s_n\frac{n+1}{2}\left(\frac{pc}{E_{\mathrm{LV,}n}}\right)^n\right],
\end{equation}
where $n=1$ or $n=2$ corresponds to linear or quadratic energy dependence of the light speed respectively, $s_n=\pm1$ indicates whether the high energy photon travels slower ($s_n=+1$) or faster ($s_n=-1$) than the low energy photon, and $E_{{\mathrm LV,}n}$ represents the $n$th-order Lorentz violation scale to be determined by the data. Taking the cosmological expansion of the universe into consideration, the light speed variation due to Lorentz violation with dispersion relation Eq.~(\ref{eq:2}) can produce a time lag (measured in the observer reference system) between two photons with different energies as~\cite{newformula,oldformula}
\begin{equation}\label{eq:3}
  \Delta t_{\mathrm{LV}}=s_n\frac{1+n}{2H_0}\frac{E^n_{\mathrm{h}}-E^n_{\mathrm{l}}}{E^n_{\mathrm{LV,}n}}\int_0^z\frac{(1+z')^n\mathrm{d}z'}
{\sqrt{\Omega_{\mathrm{m}}(1+z')^3+\Omega_{\Lambda}}},
\end{equation}
where $E_{\rm h}$ and $E_{\rm l}$ are the energies of the observed high-energy and low-energy photons,
$z$ is the redshift of the GRB source, and $H_0=\mathrm{67.3\pm 1.2 ~km s^{-1} Mpc^{-1} }$ is the present day Hubble expansion rate~\cite{pgb} while $\Omega_m=\mathrm{0.315^{+0.016}_{-0.017}}$ and $\Omega_{\Lambda}=\mathrm{0.685^{+0.017}_{-0.016}}$ are the pressureless matter density of the Universe and the dark energy density of the $\Lambda$CDM Universe respectively~\cite{pgb}.

Unfortunately, the intrinsic mechanism of GRBs is not well understood yet, hence there are big uncertainties as to how to apply Eq.~(\ref{eq:3}) to analyze the data of GRB photons. Since the photons in one GRB are not emitted from the source at the same time, the time lags observed between them consist of not only $\Delta t_{\rm LV}$ caused by the Lorentz violation as expressed in Eq.~(\ref{eq:3}), but also the intrinsic time lag at the source $\Delta t_{\rm in}$ (measured in the source reference system) between different photon events. So we should write the observed time lag between two photon events as~\cite{intrinsiclag}
\begin{equation}\label{eq:4}
  \Delta t_{\mathrm{obs}}=\Delta t_{\mathrm{LV}}+(1+z) \Delta t_{\mathrm{in}},
\end{equation}
with $\Delta t_{\mathrm{LV}}$ as in Eq.~(\ref{eq:3}).
 Then we need careful consideration for a reliable criteria to determine $\Delta t_{\mathrm{obs}}$ and $\Delta t_{\mathrm{in}}$.
 It was assumed in Refs.~\cite{zhangshu, shaolijing} that the trigger time $t_{\rm trigger}$ of Fermi Gamma-ray Burst Monitor (GBM) is the onset time of the GRB. This assumption leads to $\Delta t_{\mathrm{obs}}=t_{\rm high}-t_{\rm trigger}$, where $t_{\rm high}$ is the observed arrival time of high energy photons. In fact the trigger time of GBM is related not only with some amount of the detected low
 energy photons, but also with the detection limits of the observation equipment, i.e. GBM itself, so it may not be an objective standard adherent to the intrinsic GRB dynamics. Here we choose the peak time of the first main pulse of low energy photons in a GRB as the signal time of low energy photons, so
\begin{equation}\label{eq:Deltat}
  \Delta t_{\rm obs}=t_{\rm high}-t_{\rm low},
\end{equation}
where $t_{\rm low}$ is the peak time of the first main pulse of low energy photons. As the first low energy peak is recorded as an intensive pulse of large number of low energy photons ranging between 8$\sim $260~keV, it can serve
as a significant benchmark which is related to the intrinsic mechanism of the GRB objectively and naturally, so that
choosing it as a signal for the low energy photons is more reasonable.

For the high energy photon events from GRBs, we adopt photons that have energy higher than 10~GeV and are collected within the 90 s time window in the recent Fermi-LAT event construction~\cite{data}. Ref.~\cite{data} provided a list of high energy photons from 5 bright GRBs, i.e., GRBs 080916C, 090510, 090902B, 090926A, and 100414A. There are also later observed GRBs 130427A~\cite{zhangshu} and 140619B~\cite{140619Bz} with over 10~GeV photons and estimated redshifts. Among these GRBs, GRBs 090510 and 140619B are short bursts (with duration time less than 2~s) while others are long bursts (with duration time longer than 2~s). We list the observed energy $E_{\mathrm{obs}}$ and observed arrival time $t_{\mathrm{high}}$ of these photons in Table~\ref{tab:grbs}, where $t_{\mathrm{high}}$ is the recorded arrival time of the photon after the trigger time of GBM.

We download the GBM TTE (Time-Tagged-Events) NaI files of all GRBs listed in Table~\ref{tab:grbs} from the Fermi website~\cite{XXX} and analyze them with the RMFIT package. GBM TTE NaI files record the observed time lags (with the trigger time of GBM as the starting time point) and observed energies of a large number of photons ranging between 8$\sim$260~keV. The energy scale of these photons is so low in comparison with that of the high energy $10$~GeV photons so that the Lorentz violation effect can be neglected for these photons. We bin all these low energy events in 64~ms intervals in order to find the peak position $t_{\rm low}$, which corresponds to the first low energy peak for each GRB. The uncertainty of the determined peak position is of the order of 64 ms and is negligible in our analysis. We list the obtained results of $t_{\rm low}$ for each GRB in Table~\ref{tab:grbs}.

\begin{threeparttable}[t]
\begin{centering}
  \caption{The data of high energy photon events from GRBs with known redshifts.}
    \begin{tabular}{cccccp{23mm}<{\centering}cp{26mm}<{\centering}}
    \hline
    \hline
   GRB         & $z$           & $t_{\rm high}$~(s) & $t_{\rm low}$~(s) & $E_{\rm obs}$~(GeV) & $E_{\rm source}$~(GeV) & $\frac{\Delta t_{\rm obs}}{1+z}$~(s) & $K_{\rm 1}$ ($\times 10^{18}~\mathrm{s}~\cdot$~GeV) \\
    \hline
    080916C(1)  & $4.35\pm0.15$  & 16.545   &$5.984$   & 12.4        & 66.3              & 1.974       &  $4.46\pm 0.45 $ \\
    080916C(2)  & $4.35\pm0.15$  & 40.509  &$5.984$    & 27.4        & 146.6              & 6.453       & $9.86 \pm 0.99$ \\
    090510      & $0.903 \pm 0.003$   & 0.828  &$-0.032$     & 29.9        & 56.9            & 0.452       & $7.21 \pm 0.73 $\\
    090902B    &1.822    & 81.746   &$9.768$   & 39.9        & 112.6             & 25.506      & $12.9 \pm 1.3 $\\

        \hline
        &    & 11.671  &    & 11.9        & 33.6           & 0.674       & $3.84\pm 0.39$\\
       &    & 14.166   &   & 14.2        & 40.1             & 1.559       & $4.58 \pm 0.47$ \\
          090902Bs    & 1.822 & 26.168   & $9.768$  & 18.1        & 51.1         & 5.812       & $5.84 \pm 0.59$ \\

            &     & 42.374      &  & 12.7        & 35.8               & 11.554      &$ 4.10 \pm 0.42$\\
            &    & 45.608  &    & 15.4        & 43.5           & 12.700      & $4.97\pm 0.51$  \\
            \hline
    090926A     & $ 2.1071\pm0.0001$ & 24.835   &$4.320$   & 19.5        & 60.6              & 6.603       & $6.53\pm  0.66 $\\
    100414A     & $1.368$ & 33.365   &$0.288$   & 29.7        & 70.3                & 13.968      & $8.70\pm 0.88  $\\
    130427A     & $0.3399 \pm 0.0002$ & 18.644  &$0.544$    & 72.6        & 97.3                & 13.509      &$ 9.02 \pm 0.91 $\\
    140619B    & $2.67 \pm 0.37$   &0.613   &$0.096$    &22.7        &83.5            &0.141           &$7.96 \pm 0.82$\\
     \hline
    \hline
    \end{tabular}%

   \begin{tablenotes}
        \item Data of GRBs are from Ref.~\cite{data} (see Table~2 therein) for GRBs 080910C, 090510, 090902B, 090926A, and 100414A, from Ref.~\cite{zhangshu} (see Table~1 therein) for GRB~130427A, and from the Fermi website~\cite{XXX} for 140619B. The references for the redshifts of these GRBs are \cite{080916Cz}~(GRB~080916C), \cite{090510z}~(GRB~090510), \cite{090902Bz}~(GRB~090902B), \cite{090926Az}~(GRB~090926A), \cite{100414Az}~(GRB~100414A),  \cite{130427Az}~(GRB~130427A), and \cite{140619Bz}~(GRB 140619B). $t_{\rm high}$ and $t_{\rm low}$ denote the arrival time of the high energy photons and the peak time of the first main pulse of low energy photons respectively, with the trigger time of GBM as the zero point. Therefore $\Delta t_{\rm obs}=t_{\rm high}-t_{\rm low}$ is the observed time lag between the high energy and low energy photons.  $E_{\rm obs}$ and $E_{\rm source}$ are the energy measured by Fermi LAT and the corresponding intrinsic energy at the source of the GRBs, with the cosmological expansion factor $(1+z)$ being considered to transform $E_{\rm obs}$ to $E_{\rm source}$, i.e.,
       $E_{\rm source}=(1+z)E_{\rm obs}$. $K_{\rm 1}$ is the Lorentz violation factor calculated from Eq.~(\ref{eq:LVfactor}) with a unit of~(s~$\cdot$~GeV).
    \end{tablenotes}
  \label{tab:grbs}%
  \end{centering}
\end{threeparttable}%

\vspace*{1cm}
According to our above arguments, we take the energies of the photons observed at $t_{\rm low}$ as $E_{\rm l}$ in Eq.~(\ref{eq:3}). Therefore $E_{\rm l}$ is between 8$\sim$260~keV. Since it is extremely low compared with the energies of the high energy photon events listed in Table~\ref{tab:grbs} ($E_{\rm l}<10^{-4}E_{\rm h}$), we can set $E_{\rm l}=0$ and simplify Eq.~(\ref{eq:3}) as
\begin{equation}\label{eq:simplified}
  \Delta t_{\mathrm{LV}}=s_n\frac{1+n}{2H_0}\frac{E^n_{\mathrm{h}}}{E^n_{\mathrm{LV,}n}}\int_0^z\frac{(1+z')^n\mathrm{d}z'}
{\sqrt{\Omega_{\mathrm{m}}(1+z')^3+\Omega_{\Lambda}}}.
\end{equation}
Because we do not know $\Delta t_{\mathrm{in}}$ now, Eq.~(\ref{eq:4}) cannot be used directly. We re-express it as
\begin{equation}\label{eq:reexpress}
  \frac{\Delta t_{\mathrm{obs}}}{1+z}=s_n \frac{K_n}{E^n_{\mathrm{LV,}n}}+\Delta t_{\mathrm{in}},
\end{equation}
where $K_{\mathrm{n}}$ is the Lorentz violation factor
\begin{equation}\label{eq:LVfactor}
  K_{n}=\frac{1+n}{2H_0}\frac{E^n_{\mathrm{h}}}{1+z}\int_0^z\frac{(1+z')^n\mathrm{d}z'}
{\sqrt{\Omega_{\mathrm{m}}(1+z')^3+\Omega_{\Lambda}}}.
\end{equation}
We can find that if the energy dependence of light speed does exist, there would be a linear relation between $\Delta t_{\mathrm{obs}}/(1+z)$ and $K_{n}$. These photons with same intrinsic time lags would fall on an inclined line in the $\Delta t_{\mathrm{obs}}/(1+z)$~-~$K_{n}$ plot, and we can determine $\Delta t_{\mathrm{in}}$ of them as the intercept of the line with the $Y$ axis.

\begin{figure}
   \centering
  \includegraphics[width=110mm]{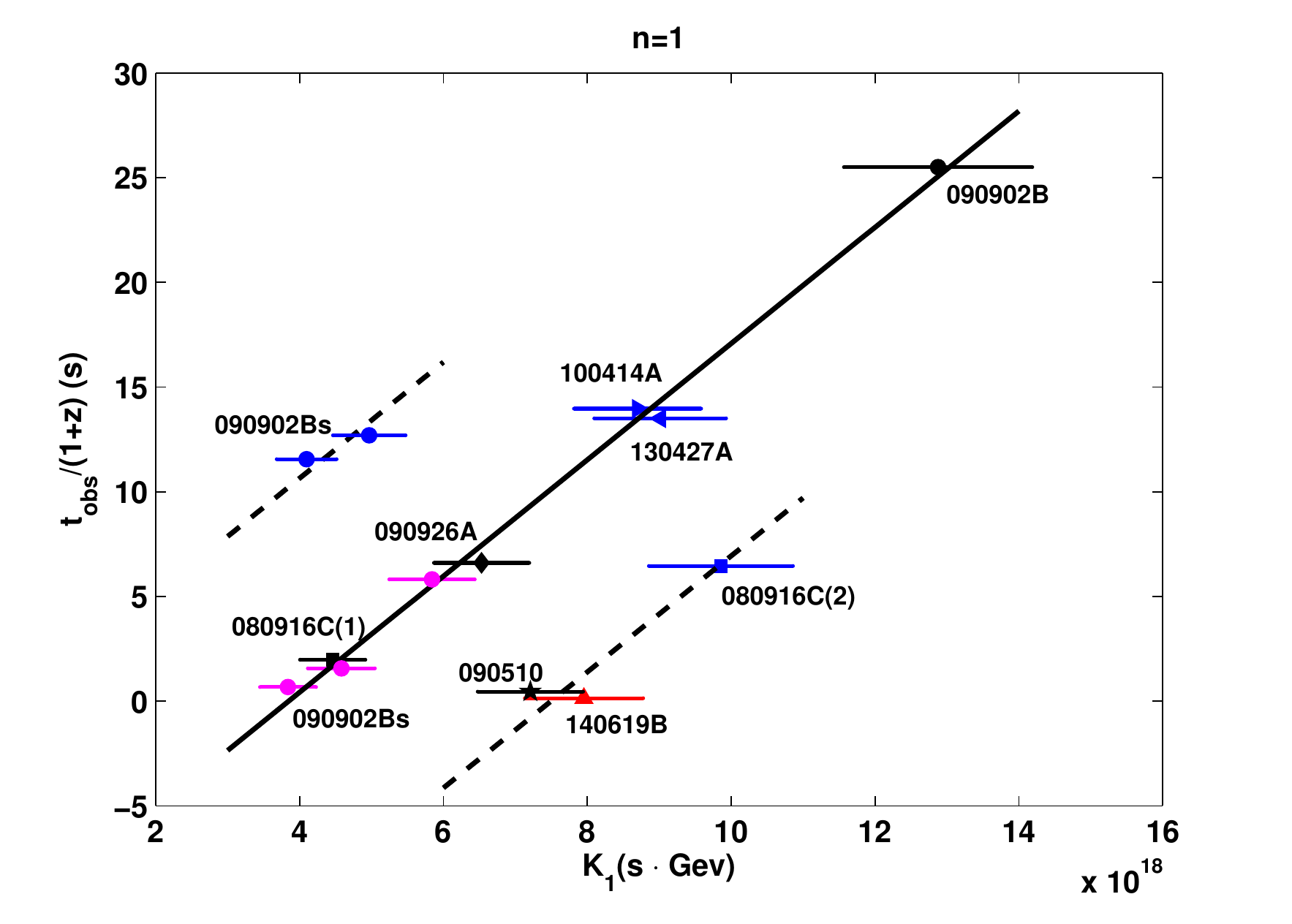}
\caption{The $\Delta t_{\rm obs}/(1+z)$ versus $K_{\rm 1}$ plot for the high energy photon events in Table~\ref{tab:grbs}.
The slopes of all the lines are $1/E_{\rm LV,1}=(2.78\pm 0.20)\times 10^{-18} ~\mathrm{GeV^{-1}}$ with $s_{1}=+1$. For the mainline with 8 events on it (black solid line), the intercept is $\Delta t_{\rm in}=-10.7\pm 1.5$~s. The lower line (dashed line) is a straight line with the same slope as that of the mainline but the intercept is fitted by the two events from short GRBs~090510 and 140619B as -20.77~s. The event of GRB~080916C(2) coincidentally falls on this line too. The upper line (dashed line) is also a straight line with the same slope but the intercept -0.47~s is fitted by two events in GRBs ~090902Bs. The standard errors of $K_1$'s are calculated with the consideration of the energy resolution of LAT~\cite{LAT} and the uncertainties of the cosmological parameters and the redshifts. Different symbols denote different GRBs. Events in black were analyzed in Ref.~\cite{shaolijing}. In Ref.~\cite{zhangshu}, events in purple and blue were added to the analysis but the purple ones were not on the mainline (see Fig. 1 in Ref.~\cite{zhangshu}). In our present paper, we add another event GRB 140619B (in red) and reanalyzed all events. Events in purple (three GRB 090902Bs on the mainline) now fall on the mainline. (For interpretation of the references to color in this figure legend, the reader is referred to the web version of this article.)}
\label{fig:figure1}
\end{figure}

We first consider the situation of the linear form correction of the light speed, i.e., the $n=1$ case. In Fig.~\ref{fig:figure1}, we draw $\Delta t_{\mathrm{obs}}/(1+z)$ versus $K_{1}$ of all the high energy photon events in Table~\ref{tab:grbs}. The $X$ axis is $K_{1}$  and the $Y$ axis is $\Delta t_{\mathrm{obs}}/(1+z)$. Strictly speaking, no assumption is made when drawing the points in Fig.~\ref{fig:figure1}. When trying to draw straight lines, we assume that some different GRBs may have similar intrinsic properties and specifically, different events may have same intrinsic time lag at the source. So these events should fall on a same line in Fig.~\ref{fig:figure1}. In fact, the points on the plot seem to be randomly distributed at first glance. But we can tell the distribution trend when noticing the event GRB 090902. Then a straight line can be drawn with 8 events from 5 long bursts on it. We consider this line as the mainline of our results. The slope of this mainline is $1/E_{\rm LV,1}=(2.78\pm 0.20)\times 10^{-18}~\mathrm{GeV^{-1}}$ and the intercept is $-10.7\pm 1.5$~s. From this point of view, we suggest that there may be a linear dispersion relation of photons at a scale of $E_{\rm LV,1}=(3.60 \pm 0.26) \times 10^{17}~ \rm GeV$. After taking the uncertainties of cosmological constants and the redshifts into consideration, the standard error of $E_{\rm LV,1}$ is $0.27\times 10^{17}~ \rm GeV$. The rest 5 events form two separated lines (dashed lines) at the two sides of the mainline. We assume that these two lines have the same slope as that of the mainline and use the data of the events on them to determine their intercepts. Two events in GRB 090902Bs fall on the upper line with an intercept determined as $-0.47$~s. By fitting the high energy photon events from two short burst GRBs 090510 and 140627B, we can determine the intercept of the lower line as -20.77~s. The event GRB 080916C(2), which has the highest intrinsic energy (146.6~GeV) of all GRB photon events, seemingly coincidentally falls on this line. We can calculate the average $E_{\rm source}$ for the events on the three lines as $40 \pm 4~\rm GeV$, $67 \pm 25~\rm GeV$ and $96 \pm 37~\rm GeV$ respectively (from up to down). It seems that photons with higher energies are more likely to be emitted from the sources earlier. This can also be considered as a support to the energy dependence of $\Delta t_{\rm in}$~\cite{zhangshu}. For a further discussion, we could fit all the points in Fig.~\ref{fig:figure1} and the slope we get is $(1.7\pm 1.4)\times 10^{-18}~\mathrm{GeV^{-1}}$.
We could also fit all the events from long bursts, then the slope we get is $(2.0\pm 1.5)\times 10^{-18}~\mathrm{GeV^{-1}}$. Obviously the uncertainties of the slopes are much bigger in the later two situations.

\begin{figure}
  \centering
  \includegraphics[width=110mm]{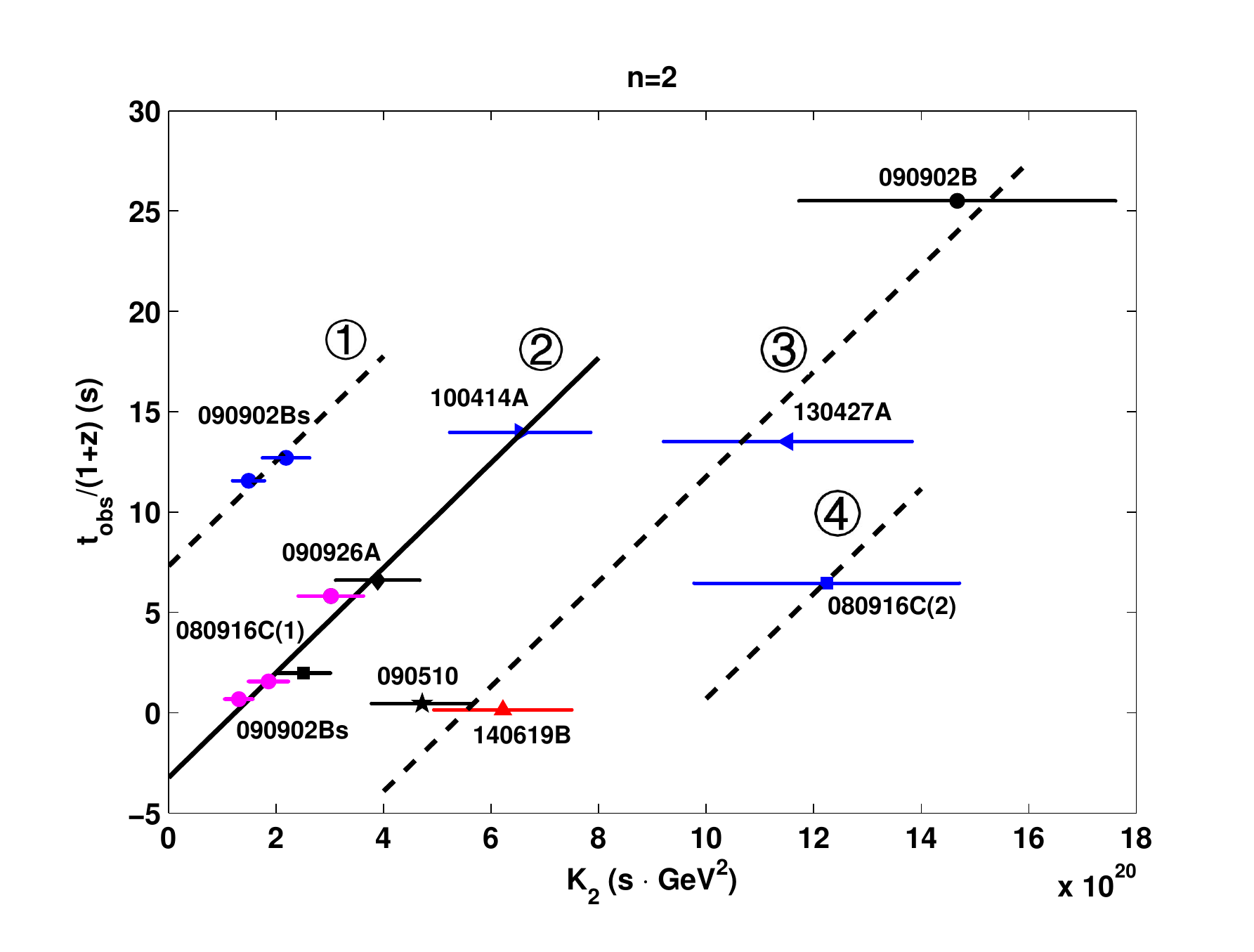}\\
  \caption{The $\Delta t_{\rm obs}/(1+z)$ versus $K_{\rm 2}$ plot for the high energy photon events in Table~\ref{tab:grbs}.
The slopes of the lines are all $1/E^2_{\rm LV,2}=(2.61\pm 0.62)\times 10^{-20}~\mathrm{GeV^{-2}}$. For the mainline $\large{\textcircled{\small{2}}}$  with 6 events on it (black solid line), the intercept is $\Delta t_{\rm in}=-3.2\pm 2.2$~s. The intercept of the upper line $ \large{\textcircled{\small{1}}}$  is 7.3~s, fitted by two events in GRB~090902Bs. The lower line $ \large{\textcircled{\small{3}}}$ is formed by the four events from GRBs~090510, 140619B, 130427A and 090902Bs and the intercept is determined to be -14.4~s. The event of 080916C(2) falls on the fourth line $ \large{\textcircled{\small{4}}}$, and its intercept is -25.4~s.}\label{fig:figure2}
\end{figure}

Then we set $n=2$ in Eqs.~(\ref{eq:3}), (\ref{eq:simplified}), (\ref{eq:reexpress}), and (\ref{eq:LVfactor}). The plot of $\Delta t_{\mathrm{obs}}/(1+z)$ versus $K_{2}$ is shown in Fig.~\ref{fig:figure2}. We still assume that all the lines should have the same slope, as what has been done for the $n=1$ case. But this time, the situation is a little different. We can see that 6 events from long GRBs form the mainline~$ \large{\textcircled{\small{2}}}$. The slope is determined to be $1/E^2_{\rm LV,2}=(2.61\pm 0.62)\times 10^{-20} ~\mathrm{GeV^{-2}}$ and this means that $E_{\rm LV,2}=(6.2\pm 0.7)\times 10^9 ~\rm GeV$.  The high energy photon events of GRBs 090902B and 130427A are no longer falling on the mainline. They form another line~$ \large{\textcircled{\small{3}}}$ together with two high energy photon events in two short GRBs~090510 and 140619B. This may indicate that these four events have same intrinsic time lag. Two events in GRB~090902Bs still form the line~$ \large{\textcircled{\small{1}}}$ above the mainline. But GRB~080916C(2) is isolated and we still draw a line~$ \large{\textcircled{\small{4}}}$ across it. This time, when we fit all the points in Fig.~\ref{fig:figure2}, the slope is $(9.8\pm 8.8)\times 10^{-21} ~\mathrm{GeV^{-2}}$. If we fit all the events from long bursts, the slope we get is $(9.8\pm 8.4)\times 10^{-21} ~\mathrm{GeV^{-2}}$.

Based on above arguments, when we compare Fig.~\ref{fig:figure1} and Fig.~\ref{fig:figure2}, it seems that there is a stronger regularity in Fig.~\ref{fig:figure1} with 8 events among 13
to fall dramatically on a straight line (mainline), whereas there are only 6 events to fall on a straight line in Fig.~\ref{fig:figure2}. Moreover, the only two events of short GRBs
and the event with the highest intrinsic energy photon also fall on a straight line parallel to the mainline in Fig.~\ref{fig:figure1}, whereas they fall on two different lines in Fig.~\ref{fig:figure2}.
So Fig.~\ref{fig:figure1} is more favored with stronger regularities and we need to put more attention
on this figure. In Fig.~\ref{fig:figure1}, the 8 high energy photon events falling on the mainline are from 5 different long GRBs and 4 events among them are from an identical GRB~090902B. This indicates that they have the same intrinsic lag if the light speed energy variation does exist. This regularity may not just be a coincidence and we think it is a support for the existence of a linear form of light speed energy dependence at a scale of $E_{\rm LV,1}=(3.60\pm 0.26)\times 10^{17}~\rm GeV$. Another two events from GRB~090902B fall on the upper line. This means that they have bigger intrinsic time lags than the rest four events which fall on the mainline in the same GRB~090902B. This is reasonable because different photons in a GRB may be emitted from the source at different times.
Furthermore, two high energy photon events from the short GRBs 090510 and 140619B can be satisfactorily fitted by a line with the same slope as the mainline. We think this corroborates the assumption that short GRBs may have similar intrinsic, though not well not understood yet, mechanism.
We can also see that the photon event with  highest intrinsic energy, GRB 080916C(2), falls on the lower line formed by events from short GRBs. If this is not only a coincidence, it may mean that photon events with ultra-high intrinsic energies in long bursts have different intrinsic emission mechanism in comparison with that of a few ten~GeV scale photons.
Finally, in our analysis, photons with high energy are emitted at the source of GRBs before the emission of low energy photons. However, what we observe is that high energy photons arrive later than low energy photons. It is possible that the delay of high energy photons is caused by LV. If future observation and more data support our analysis, we may have revealed some intriguing intrinsic properties of the GRBs, and some previous analysis about GRBs shall be reconsidered.

We can compare our work with that in Ref.~\cite{zhangshu}, where the trigger time of GBM was adopted as the onset time of the GRB. It was shown that 5 events from 5 different GRBs fall on the mainline in Fig.~1 therein. In contrast, in the present work we choose the peak time of the first main pulse of low energy photons as the time benchmark of low energy photons. It is a surprise that three more high energy photon events in GRB~090902B fall on the mainline now, so that we have 8 events among 13 events to fall on a same line. We thus reveal a stronger regularity as shown in Fig.~\ref{fig:figure1} for the high energy photon events listed in Table~\ref{tab:grbs}.
In addition, we add one more high energy photon event from the short GRB~140619B to fall on the short GRB line as speculated in Refs.~\cite{zhangshu,shaolijing}. These new results provide us with suggestive regularities
of the high energy photon events from long and short GRBs.

The regularities we revealed in Fig.~\ref{fig:figure1} suggest a linear form energy dependence of light speed at a scale of $E_{\rm LV,1}=(3.60\pm 0.26)\times 10^{17}~\rm GeV$. Such result can be understood
with two possible explanations. The first is that such light speed energy dependence is due to the Lorentz invariance violation, so that we reveal an evidence for the Lorentz violation. It is also possible that
the matter in the universe leads to a dispersion as expressed in Eq.~(\ref{eq:1}), so the regularities indicate the consequence of matter effect on light speed in the universe~\cite{mattereffect}.

\begin{figure}[!h]
   \centering
  \includegraphics[width=110mm]{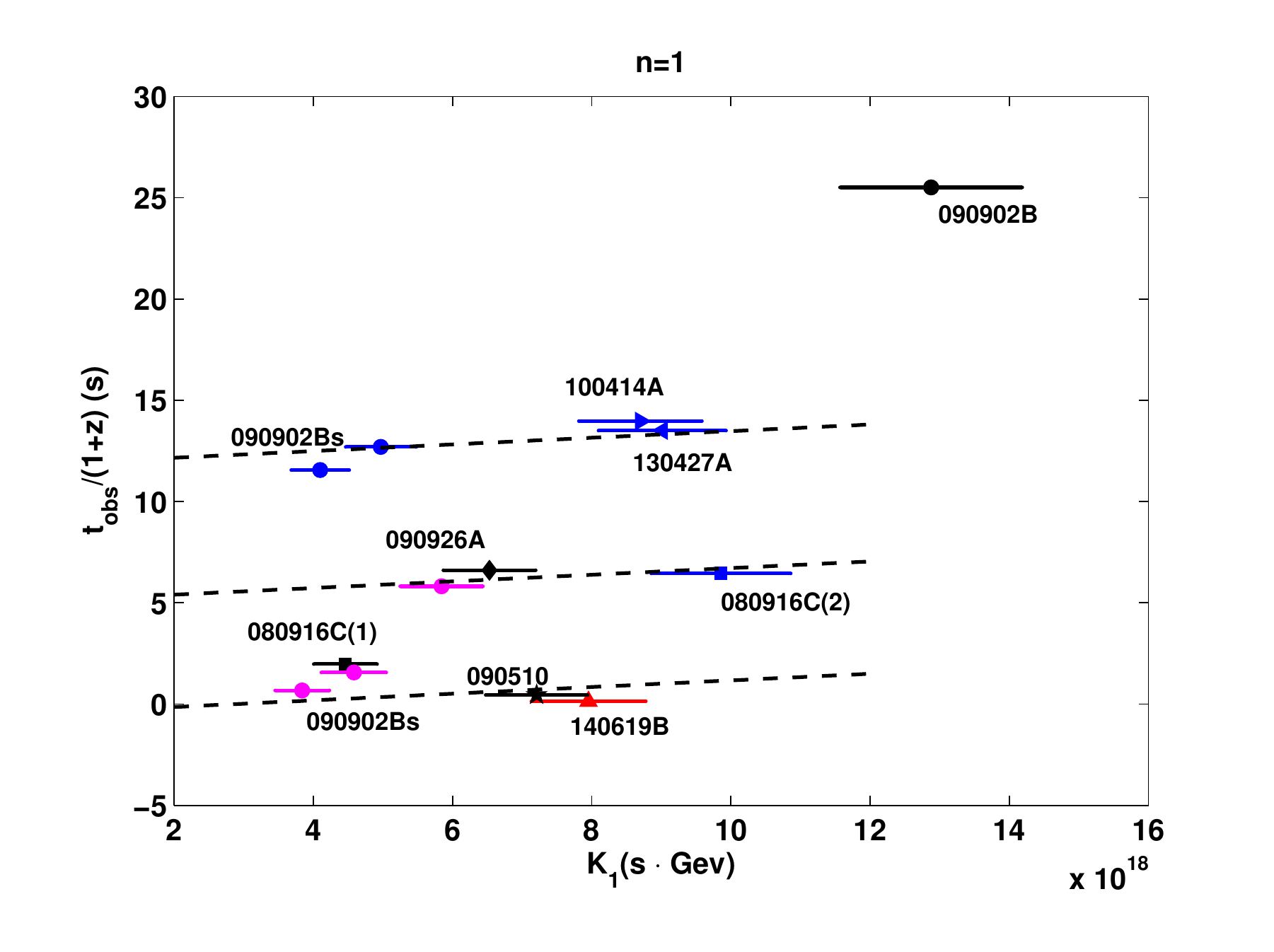}
\caption{The $\Delta t_{\rm obs}/(1+z)$ versus $K_{\rm 1}$ plot for the high energy photon events in Table~\ref{tab:grbs}. The data are fitted in a different way from that in Fig.~\ref{fig:figure1} . The slopes of all the lines are $1/E_{\rm LV,1}=(1.64\pm 0.30)\times 10^{-19} ~\mathrm{GeV^{-1}}$ with $s_{1}=+1$. The intercepts of the three lines are +11.83s, +5.07s, -0.47~s respectively.}
%For the mainline with 8 events on it (black solid line), the intercept is $\Delta t_{\rm in}=-10.7\pm 1.5$~s. The lower line (dashed line) is a straight line with the same slope as that of the mainline but the intercept is fitted by the two events from short GRBs~090510 and 140619B as -20.77~s. The event of GRB~080916C(2) coincidentally falls on this line too. The upper line (dashed line) is also a straight line with the same slope but the intercept -0.47~s is fitted by two events in GRBs ~090902Bs. The standard errors of $K_1$'s are calculated with the consideration of the energy resolution of LAT~\cite{LAT} and the uncertainties of the cosmological parameters and the redshifts.%}
\label{fig:figure3}
\end{figure}

In fact, a number of groups have used GRB data to study the light speed variation. Ellis et al.~\cite{intrinsiclag} used the weighted averages of the time-lags calculated using correlated features in the GRB light curves
from the HETE and SWIFT data and assumed that the intrinsic time lag is the same for all GRBs, i.e., fitted all the data on a single line. They suggested a statistically robust lower limit $E_{\mathrm{LV}}>1.4 \times 10^{16}~\rm GeV$. If we also fit all the points in Fig.~\ref{fig:figure1} on a single line, the result is $E_{\rm LV}=5.9 \times 10^{17}~\rm GeV$. Chang et al.~\cite{changzhe} used the continuous spectra of 20 short GRBs detected by the Swift satellite and gave a conservative lower limit of quantum gravity energy scale as $E_{\rm LV}>5.05 \times 10^{14}~\rm GeV$. Vasileiou et al.~\cite{Vlasios Vasileiou} used high-energy observations from the LAT data of gamma-ray burst GRB090510 and tested a model in which photon speeds are distributed normally around $c$ with a standard deviation proportional to the photon energy. They constrained the characteristic energy scale of the model as $E_{\rm LV}>3.4 \times 10^{19}~\rm GeV$. In comparison with previous results, our result seems to be compatible with some but also different from others. One main reason for the difference is that there are actually big randomicities in the analyzes due to different assumptions. For example, in our analysis we could divide the data into different groups and then fit them on different parallel lines. One example is shown in Fig.~\ref{fig:figure3}, where we divide the data into 4 different groups with each containing  1-5 GRB events. Then we get $E_{\rm LV}=(6.1\pm1.1) \times 10^{18}~\rm GeV$, which is different from that in Fig.~\ref{fig:figure1}. But we can see that the regularities in Fig.~\ref{fig:figure1} are more remarkable, with 8 out of 13 events to fall on a same line. Therefore we still need to check whether such regularity can still persist or not with more energetic GRB events in the future.

In conclusion, by analysing the data of high energy photon events from seven GRBs with known redshifts, we find that eight events from five long GRBs fall on an inclined line in the $\Delta t_{\mathrm{obs}}/(1+z)$ versus $K_{1}$ plot, i.e., Fig.~\ref{fig:figure1}. We find also that the two events from two short GRBs and the rare event of GRB 080916C(2) with highest intrinsic energy at the source form a line in parallel with the above line. If such results are not due to statistical fluctuation, we consider these fascinating regularities as an indication for a linear form modification of light speed $v(E)=c(1-E/E_{\mathrm{LV}})$, where $E$ is the photon energy and $E_{\mathrm{LV}}=(3.60 \pm 0.26) \times 10^{17}$~GeV serves as the Lorentz violation scale. From a conservative viewpoint, we may consider the results of our work as a suggestion for a lower limit of $E_{\mathrm{LV}} \gtrsim 10^{17}$~GeV.

%\begin{acknowledgements}

{\bf Acknowledgements}\quad \quad \quad
This work is supported by National Natural Science Foundation of China (Grants Nos.~11120101004 and 11475006) and the National Fund for Fostering Talents of Basic Science (Grant Nos.~J1103205 and J1103206). It is also supported by the Undergraduate Research Fund of Education Foundation of Peking University.
%\end{acknowledgements}





\end{document}